\documentclass[prb,preprint,onecolumn,showpacs,preprintnumbers,amsmath,amssymb,superscriptaddress,floatfix]{revtex4-2}
\usepackage[T1]{fontenc}
\usepackage[ansinew]{inputenc}
\usepackage{graphics}
\usepackage{dcolumn}
\usepackage{bm}
\usepackage{amsthm}
\usepackage{amsmath}
\usepackage{amssymb}
\usepackage{diagbox}
\usepackage{hyperref}
\usepackage{url}
\usepackage{xcolor}

\def\afflux{Department of Physics and Materials Science, University of Luxembourg, 162A~Avenue de la Faiencerie, L-1511 Luxembourg, Grand Duchy of Luxembourg}
\def\affbrin{Research Center for Quantum Physics, National Research and Innovation Agency, South Tangerang, Indonesia}

\begin{document}

\title{Neutron-scattering signature of the Dzyaloshinskii-Moriya interaction in nanoparticles}

\author{Evelyn Pratami Sinaga}\email{evelyn.sinaga@uni.lu}
\address{\afflux}
\author{Michael P.\ Adams}
\address{\afflux}
\author{Eddwi H.\ Hasdeo}
\address{\afflux}
\address{\affbrin}
\author{Andreas Michels}\email{andreas.michels@uni.lu}
\address{\afflux}

\date{\today}

\begin{abstract}
The antisymmetric Dzyaloshinkii-Moriya interaction (DMI) arises in systems with broken inversion symmetry and strong spin-orbit coupling. In conjunction with the isotropic and symmetric exchange interaction, magnetic anisotropy, the dipolar interaction, and an externally applied magnetic field, the DMI supports and stabilizes the formation of various kinds of complex mesoscale magnetization configurations, such as helices, spin spirals, skyrmions, or hopfions. A question of importance in this context addresses the neutron-scattering signature of the DMI, in particular in nanoparticle assemblies, where the related magnetic scattering signal is diffuse in character and not of the single-crystal diffraction-peak-type, as it is e.g.\ seen in the B20 compounds. Using micromagnetic simulations we study the effect of the DMI in spherical FeGe nanoparticles on the {\it randomly-averaged} magnetic neutron scattering observables, more specifically on the spin-flip small-angle neutron scattering cross section, the related chiral function, and the pair-distance distribution function. Within the studied parameter space for the particle size ($60 \, \mathrm{nm} \leq L \leq 200 \, \mathrm{nm}$) and the applied magnetic field ($-1 \, \mathrm{T} \leq \mu_0 H_0 \leq 1 \, \mathrm{T}$), we find that the chiral function is only nonzero when the DMI is taken into account in the simulations. This result is discussed within the context of the symmetry properties of the magnetization Fourier components and of the involved energies under space inversion. Finally, for small applied magnetic fields, we provide an easy-to-implement analytical correlation function for the DMI-induced spin modulations (with wave vector $k_{\mathrm{d}}$). The corresponding randomly-averaged spin-flip SANS cross section reproduces the main features found in the numerical simulations.
\end{abstract}

\date{\today}

\maketitle

\section{Introduction}
\label{introduction}

The Dzyaloshinkii-Moriya interaction (DMI) is due to the relativistic spin-orbit coupling and arises in condensed-matter systems that exhibit a crystal-field environment with no inversion symmetry~\cite{dzya58,moriya60}. This is e.g.\ the case in noncentrosymmetric crystal structures (such as MnSi or FeGe), where the DMI is intrinsic to the material~\cite{bogdanov89,bogdanov94}, or in microstructural-defect-rich samples (such as ultrathin film architectures, mechanically-deformed magnets, or magnetic nanoparticles), where the DMI is due to the breaking of structural inversion symmetry at the defect sites~\cite{arrott1963,fert1980,Fedorov1997,lott08,faehnle2010,butenko2013,michelsPRB2016,beyerlein2018,ono2019,michelsdmi2019}. The recent renaissance of the DMI is largely related to the fact that it is the essential ingredient for the stabilization of various types of topological spin structures, such as skyrmions, which might be of importance for spintronics applications (see, e.g., Refs.~\cite{pflei2009,Pfleiderer_SkL_rev,Nagaosa2013,wiesendanger2016,sky_Roadmap2020,bogdanov2020} and references therein).

A question of interest addresses the signature of the DMI in experimental data. This is a highly nontrivial point since the DMI generally appears only in concert with other, usually much stronger, magnetic interactions, such as the isotropic and symmetric exchange interaction or the magnetodipolar interaction; these may then mask the fingerprint of the DMI in a particular measurement. In contrast to isotropic exchange, which favors the parallelism of magnetic moments, the DMI energy prefers noncollinear spin configurations, similar to the magnetostatic interaction that gives rise to flux-closure patterns~\cite{perigo2014,metlov2021,bersweilerprb2023}. Measurement of the topological Hall effect is frequently used to establish the occurrence of chiral spin structures (where the DMI plays an important role), but a recent review critically discusses the challenges and limitations of this method~\cite{THE2023}. Likewise, Lorentz transmission electron microscopy also allows the study of topological spin structures, as was recently shown for the case of hopfion rings in an FeGe crystal~\cite{blugel2023}. Magnetic neutron scattering is another important technique in this regard since the cross section for polarized neutrons contains the so-called chiral function; for instance, using an advanced polarized diffuse neutron diffraction technique, Schweika~{\it et al.}~\cite{schweika2022} have experimentally discovered a chiral spin liquid ground state in a single-domain single crystal of the noncentrosymmetric swedenborgite compound $\mathrm{YBaCo_3FeO_7}$.

Here, we focus on magnetic small-angle neutron scattering (SANS), which is a powerful method for the investigation of mesoscale spin structures within the volume of magnetic media~\cite{rmp2019,michelsbook}. The relevant quantity for understanding magnetic SANS is the three-dimensional magnetization vector field $\mathbf{M} = \mathbf{M}(\mathbf{r})$, which can be computed using the continuum theory of micromagnetics~\cite{brown}. The Fourier transform $\widetilde{\mathbf{M}} = \widetilde{\mathbf{M}}(\mathbf{q})$ of the real-space spin structure then determines the magnetic neutron scattering cross section. Using numerical micromagnetic computations, we study the signature of the DMI in spherical FeGe nanoparticles in the {\it randomly-averaged} SANS observables, in particular in the spin-flip small-angle neutron scattering cross section and the related chiral function, which can be obtained from polarized SANS measurements via an uniaxial polarization analysis~\cite{michels2010epjb,kryckaprl2014,zakutna2020}. As we will see, the chiral function in the polarized SANS cross section is a very important means to disentangle the presence of the DMI in nanoparticles.

The article is organized as follows:~In Sec.~\ref{mumagdetails} we provide information on the micromagnetic simulations, we display the expressions for the spin-flip SANS cross section, the chiral function, and for the pair-distance distribution function, and we recall the basic symmetry properties of these quantities. In Sec.~\ref{ressec} we present and discuss the simulation results, while Sec.~\ref{summary} summarizes the main findings of this study and provides an outlook on future challenges. The Appendix displays additional results for the randomly-averaged magnetization curve, the spin-flip SANS cross section, the chiral function, and for the pair-distance distribution function of FeGe nanoparticles.

\section{Details on the micromagnetic simulations, spin-flip SANS cross section, chiral function, and pair-distance distribution function}
\label{mumagdetails}

\begin{figure}[tb!]
\centering
\resizebox{0.35\columnwidth}{!}{\includegraphics{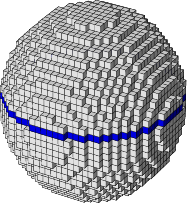}}
\caption{Illustration of the discretization of a nanosphere into cubical cells with a size of $2 \times 2 \times 2 \, \mathrm{nm^3}$. The blue-colored cells mark the middle layer through the center of the particle for which the topological charge [Eq.~(\ref{topocharge:eq})] has been computed. The small-angle scattering from such a sphere in the saturated state agrees very well with the analytical solution for the sphere form factor~\cite{laura2020,evelynprb2023}.}
\label{fig1}
\end{figure}

We were using the open-source software package MuMax3 (version~3.10) for the micromagnetic simulations~\cite{mumax3new,leliaert2019}. This progam is a widely-used micromagnetic simulation tool that enables researchers to investigate the static and dynamic nanoscale behavior of magnetic materials. Mumax3 employs a finite-difference discretization scheme of space using an orthorhombic grid of cells (see Fig.~\ref{fig1}). The following contributions to the total magnetic Gibbs free energy $G = E_{\mathrm{z}} + E_{\mathrm{d}} + E_{\mathrm{ani}} +E_{\mathrm{ex}} + E_{\mathrm{dmi}}$ were taken into account:~Zeeman energy $E_{\mathrm{z}}$ in the external magnetic field $\mathbf{H}_0$, dipolar (magnetostatic) interaction energy $E_{\mathrm{d}}$, energy of the (cubic) magnetocrystalline anisotropy $E_{\mathrm{ani}}$, isotropic and symmetric exchange energy $E_{\mathrm{ex}}$, and the Dzyaloshinkii-Moriya interaction (DMI) energy $E_{\mathrm{dmi}}$. The continuum expressions for these energies are the following~\cite{brown}:
\begin{eqnarray}
\label{eq1}
E_{\mathrm{z}} &=& - \mu_0 M_{\mathrm{s}} \int \mathbf{m} \cdot \mathbf{H}_0 \, dV , \\
\label{eq2}
E_{\mathrm{d}} &=& - \frac{1}{2} \mu_0 M_{\mathrm{s}} \int \mathbf{m} \cdot \mathbf{H}_{\mathrm{d}} \, dV , \\
\label{eq3}
E_{\mathrm{ani}} &=& K_{\mathrm{c}1} \int \big[ (\mathbf{c}_1 \cdot \mathbf{m})^2 (\mathbf{c}_2 \cdot \mathbf{m})^2 + (\mathbf{c}_1 \cdot \mathbf{m})^2 (\mathbf{c}_3 \cdot \mathbf{m})^2 \\
&& + (\mathbf{c}_2 \cdot \mathbf{m})^2 (\mathbf{c}_3 \cdot \mathbf{m})^2 \big] \, dV , \nonumber \\
\label{eq4}
E_{\mathrm{ex}} &=& A \int \left[ (\nabla m_x)^2 + (\nabla m_y)^2 + (\nabla m_z)^2 \right] \, dV , \\
\label{eq5}
E_{\rm dmi}&=& D \int \mathbf{m}\cdot \nabla \times \mathbf{m} \, dV ,
\end{eqnarray}
where $\mu_0 = 4\pi \times 10^{-7} \, \mathrm{Tm/A}$, $\mathbf{m}(\mathbf{r}) = \mathbf{M}(\mathbf{r})/M_{\mathrm{s}}$ denotes the unit magnetization vector field with $M_{\mathrm{s}}$ being the saturation magnetization, $\mathbf{H}_0$ is the (constant) applied magnetic field, $\mathbf{H}_{\mathrm{d}}(\mathbf{r}; \mathbf{M}(\mathbf{r}))$ is the magnetostatic self-interaction field, $K_{\mathrm{c}1}$ is the first-order cubic anisotropy constant with the $\mathbf{c}_{1,2,3}$~vectors representing the local (mutually perpendicular) cubic anisotropy axes, $A$ is the exchange-stiffness constant, $D$ is the bulk DMI constant, and the integrals are taken over the volume of the sample. In the simulations, we used the following material parameters for FeGe~\cite{takagi2017,hertel2021}:~$M_{\mathrm{s}} = 384 \, \mathrm{kA/m}$, $K_{\mathrm{c}1} = 1.0 \times 10^{4}\,\mathrm{J/m^3}$, $A = 8.8 \times 10^{-12} \, \mathrm{J/m}$, and $D = 1.6 \times 10^{-3}\, \mathrm{J/m^{-2}}$. These values result in a magnetostatic exchange length of $l_{\mathrm{s}} = \sqrt{2A/(\mu_0 M_{\mathrm{s}}^2)} = 9.7 \, \mathrm{nm}$, a domain-wall parameter of $l_{\mathrm{k}} = \sqrt{A/K_{\mathrm{c}1}} = 29.7 \, \mathrm{nm}$, and in a helical period of $l_{\mathrm{d}} = 4\pi A/D = 69.1 \, \mathrm{nm}$~\cite{kronparkinhandbook07,hertel2021}. We refer to Ref.~\cite{mumax3new} for a discussion of how the above continuum expressions for the magnetic energies are numerically implemented on a discrete spatial grid.

By noting that the magnetic field and the magnetization are both pseudovectors that exhibit an even behavior under the space-inversion operation ($\mathbf{r} \rightarrow -\mathbf{r}$)~\cite{jacksonen}, it is seen that Eqs.~(\ref{eq1})$-$(\ref{eq4}) are invariant under the parity transformation. On the other hand, due to the fact that the del operator $\nabla$ breaks the space-inversion symmetry, the DMI energy [Eq.~(\ref{eq5})] is a pseudoscalar that acquires a minus sign on $\mathbf{r} \rightarrow -\mathbf{r}$; in other words, the DMI energetically favors a particular chirality in the system, which would otherwise be chirally-symmetric. These symmetry properties remain after the variation of the total magnetic Gibbs free energy (with respect to $\mathbf{m}$) is carried out to obtain the partial differential equations that describe the system behavior. For the static case, the ensuing equations for the equilibrium magnetization configuration (Brown's equations) can be conveniently written in the form of a torque equation, $\mathbf{m}(\mathbf{r}) \times \mathbf{h}_{\mathrm{eff}}(\mathbf{r}) = \mathbf{0}$, where $\mathbf{h}_{\mathrm{eff}} = -\mu_0^{-1} M_{\mathrm{s}}^{-2} \delta G / \delta \mathbf{m} = \mathbf{H}_{\mathrm{eff}}/M_{\mathrm{s}}$ denotes the (dimensionless) effective magnetic field~\cite{brown}. More specifically, for the energies Eqs.~(\ref{eq1})$-$(\ref{eq5}), the effective field reads:
\begin{equation}
\label{heff:eq}
\mathbf{h}_{\mathrm{eff}} = \mathbf{h}_0 + \mathbf{h}_{\mathrm{d}} + \mathbf{h}_{\mathrm{ani}} + \mathbf{h}_{\mathrm{ex}} + \mathbf{h}_{\mathrm{dmi}} ,
\end{equation}
where $\mathbf{h}_0 = \mathbf{H}_0/M_{\mathrm{s}}$ is the normalized applied magnetic field, $\mathbf{h}_{\mathrm{d}} = \mathbf{H}_{\mathrm{d}}/M_{\mathrm{s}}$ is the magnetostatic field,
\begin{eqnarray}
\mathbf{h}_\mathrm{ani} &=& - \frac{2 K_{\mathrm{c}1}}{\mu_0 M^2_{\mathrm{s}}} \big\{ \mathbf{c}_1 (\mathbf{c}_1 \cdot \mathbf{m}) \left[ (\mathbf{c}_2 \cdot \mathbf{m})^2 + (\mathbf{c}_3 \cdot \mathbf{m})^2 \right] \\
&& + \mathbf{c}_2 (\mathbf{c}_2 \cdot \mathbf{m}) \left[ (\mathbf{c}_1 \cdot \mathbf{m})^2 + (\mathbf{c}_3 \cdot \mathbf{m})^2 \right] \nonumber \\
&& + \mathbf{c}_3 (\mathbf{c}_3 \cdot \mathbf{m}) \left[ (\mathbf{c}_1 \cdot \mathbf{m})^2 + (\mathbf{c}_2 \cdot \mathbf{m})^2 \right] \big\} \nonumber
\end{eqnarray}
represents the cubic anisotropy field, $\mathbf{h}_{\mathrm{ex}} = l_{\mathrm{s}}^2 \nabla^2 \mathbf{m} = l_{\mathrm{s}}^2 \{ \nabla^2 m_x, \nabla^2 m_y, \nabla^2 m_z \}$ is the exchange field, and $\mathbf{h}_{\mathrm{dmi}} = - l_{\mathrm{dmi}} \nabla \times \mathbf{m}$ denotes the conjugate field related to the DMI ($l_{\mathrm{dmi}} = 2D/(\mu_0 M_{\mathrm{s}}^2)$, $l_{\mathrm{dmi}} = 17.3 \, \mathrm{nm}$ for FeGe). On space inversion, only $\mathbf{h}_{\mathrm{dmi}}$ changes its sign.

Figure~\ref{fig1} displays the structural model used in the micromagnetic SANS simulations of FeGe nanospheres. We carried out simulations for sphere diameters $L$ ranging between $60 \, \mathrm{nm} \leq L \leq 200  \, \mathrm{nm} $. The sphere volume was discretized into cubical cells ``$i$'' with a size (volume) of $V_i = 2 \times 2 \times 2 \ \, \mathrm{nm}^3$ (finite-difference method). This cell size is motivated by the above values for $l_{\mathrm{s}}$, $l_{\mathrm{k}}$, and $l_{\mathrm{d}}$ and by the aim to resolve spatial variations in the magnetization that are smaller than these characteristic length scales (see the discussion in Refs.~\cite{Lopez-Diaz_2012,Leliaert_2018}). In each cell ``$i$'' with volume $V_i$, the magnetic moment vector is given by $\boldsymbol{\mu}_i = \boldsymbol{\mu}_i(\mathbf{r}) = M_{\mathrm{s}} V_i \mathbf{m}_i(\mathbf{r})$, where $\mathbf{m}_i(\mathbf{r})$ is a unit vector along the local direction of the magnetization. Open boundary conditions were used, since we are interested in the scattering behavior of an ensemble of noninteracting single particles having random easy-axis orientations $\mathbf{c}_{1,2,3}$.\footnote{In the simulations, $\mathbf{c}_1$ is a random unit vector that is generated using two random angles. A second random unit vector, say $\mathbf{a}$, is generated by another set of random angles, such that $\mathbf{c}_2 = (\mathbf{c}_1 \times \mathbf{a})/|\mathbf{c}_1 \times \mathbf{a}|$ and $\mathbf{c}_3 = \mathbf{c}_1 \times \mathbf{c}_2$.} All simulations were carried out by first saturating the nanoparticle by a strong external field $\mathbf{H}_0$, and then the field was decreased in steps of typically $5 \, \mathrm{mT}$ following the major hysteresis loop. For each step of $H_0$ and for each particular easy-axis orientation, we have obtained the equilibrium spin structure $m_{x,y,z}(x,y,z)$ by employing both the ``Relax'' and ``Minimize'' functions of MuMax3. The former solves the Landau-Lifshitz-Gilbert equation without the precessional term and the latter uses the conjugate-gradient method to find the configuration of minimum energy. To obtain an idea on the existence of a possible skyrmion texture in the FeGe nanoparticles, we have numerically computed the topological charge $Q$ for the middle layer within the $x$-$y$~plane according to (compare Fig.~\ref{fig1})~\cite{braun2012,mulkers2020}:
\begin{equation}
\label{topocharge:eq}
Q = \frac{1}{4\pi} \int \mathbf{m} \cdot \left( \frac{\bf \partial m}{\partial x} \times \frac{\bf \partial m}{\partial y} \right) dx dy .
\end{equation}
For an idealized vortex-type planar structure with $\mathbf{m} = \frac{1}{2} \{-y, x, 0 \}$ and $\nabla \times \mathbf{m} = \{0, 0, 1 \}$, one finds $Q = 0$, while $Q = \pm 1$ for skyrmions~\cite{Nagaosa2013}. However, one should keep in mind that the latter values suppose that the skyrmion fully fits inside the particle and that the magnetization vector far away from the skyrmion center approaches a constant value (so-called ferromagnetic background). This is of course fulfilled by the mathematical trial functions that are used to describe N\'{e}el and Bloch skyrmions (e.g., \cite{bogdanov94}). Here, for finite-sized nanoparticles, the magnetodipolar interaction (which is always present) aims to avoid volume and surface charges by demanding that $\nabla \cdot \mathbf{m} = 0$ and $\mathbf{m} \cdot \mathbf{n} = 0$, where $\mathbf{n}$ denotes the local unit normal vector to the surface. This implies that the surface spins (those far away from the skyrmion center, which is supposed to be localized in the sphere center due to symmetry reasons) may not attain a constant value, but vary over the sphere surface. Therefore, in micromagnetic simulations using open boundary conditions on finite-sized systems one should not expect to find $Q$~values very close to unity.

\begin{figure}[tb!]
\centering
\resizebox{0.70\columnwidth}{!}{\includegraphics{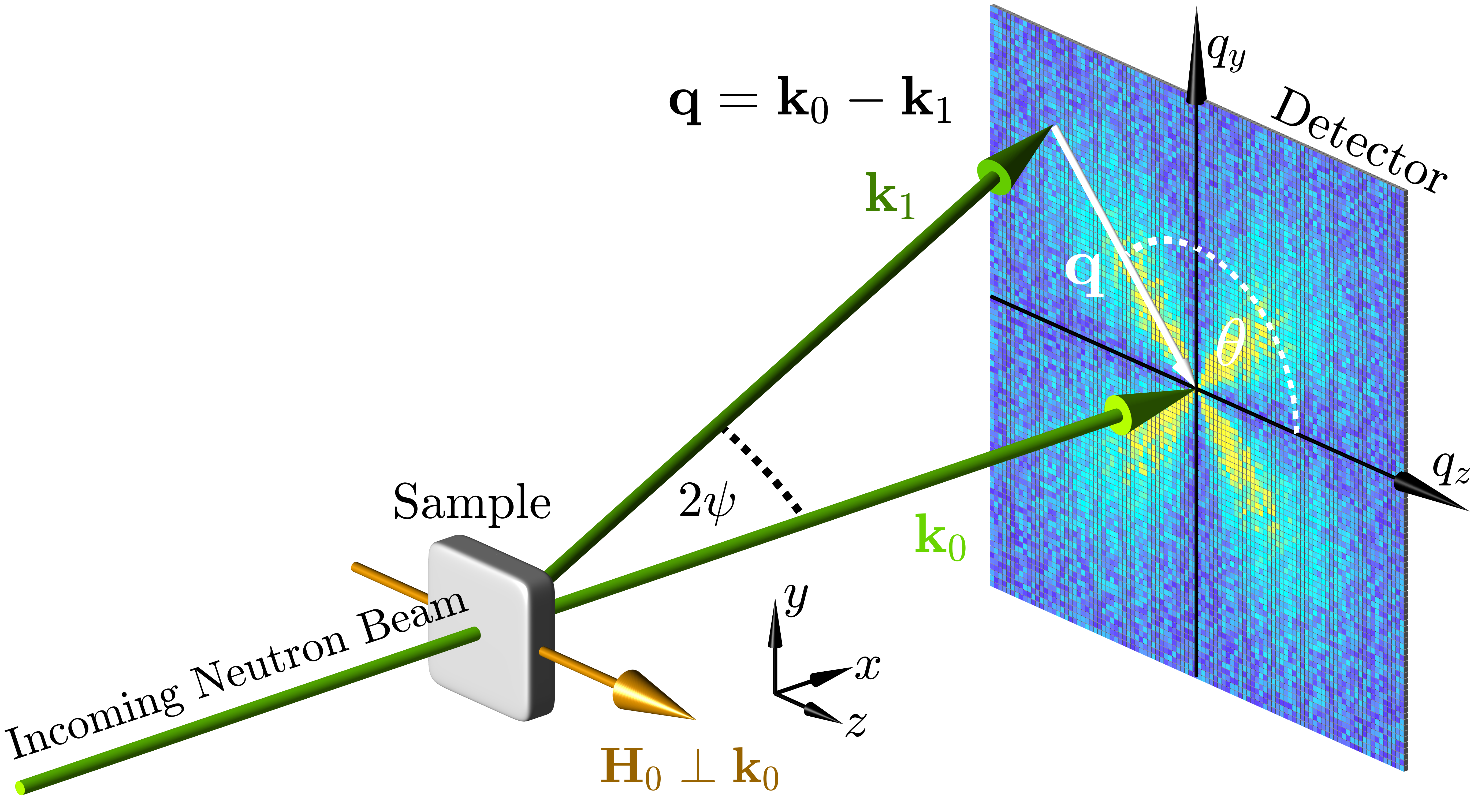}}
\caption{Sketch of the scattering geometry assumed in the micromagnetic simulations. The neutron optical elements (polarizer, spin flipper, analyzer) that are required to measure the spin-flip SANS cross section and the chiral function are not drawn. The applied magnetic field $\mathbf{H}_0 \parallel \mathbf{e}_z$ is perpendicular to the wave vector $\mathbf{k}_0 \parallel \mathbf{e}_x$ of the incident neutron beam ($\mathbf{H}_0 \perp \mathbf{k}_0$). The momentum-transfer or scattering vector $\mathbf{q}$ is defined as the difference between $\mathbf{k}_0$ and $\mathbf{k}_1$, i.e., $\mathbf{q} = \mathbf{k}_0 - \mathbf{k}_1$. SANS is usually implemented as elastic scattering ($k_0 = k_1 = 2\pi / \lambda$), and the component of $\mathbf{q}$ along the incident neutron beam, here $q_x$, is much smaller than the other two components so that $\mathbf{q} \cong \{ 0, q_y, q_z \} = q \{ 0, \sin\theta, \cos\theta \}$. This demonstrates that SANS probes predominantly correlations in the plane perpendicular to the incident beam. The angle $\theta = \angle(\mathbf{q}, \mathbf{H}_0)$ is used to describe the angular anisotropy of the recorded scattering pattern on the two-dimensional position-sensitive detector. For elastic scattering, the magnitude of $\mathbf{q}$ is given by $q = (4\pi / \lambda) \sin(\psi/2)$, where $\lambda$ denotes the mean wavelength of the neutrons and $\psi$ is the scattering angle.}
\label{fig2}
\end{figure}

The quantities of interest are the elastic differential spin-flip scattering cross section and the related so-called chiral function, which are usually obtained in an uniaxial polarization-analysis experiment~\cite{michels2010epjb,kryckaprl2014,zakutna2020}. For the most commonly used scattering geometry in magnetic SANS experiments, where the applied magnetic field $\mathbf{H}_0 \parallel \mathbf{e}_z$ is perpendicular to the wave vector $\mathbf{k}_0 \parallel \mathbf{e}_x$ of the incident neutrons (see Fig.~\ref{fig2}), the two spin-flip SANS cross sections $d \Sigma^{+-}_{\mathrm{sf}} / d \Omega$ and $d \Sigma^{-+}_{\mathrm{sf}} / d \Omega$ can be written as~\cite{rmp2019,michelsbook}:
\begin{eqnarray}
 \label{eq:equation1a}
\frac{d \Sigma^{+-}_{\mathrm{sf}}}{d \Omega} &=& \frac{8 \pi^3}{V} b_{\mathrm{H}}^2 \left( |\widetilde{M}_x|^2 + |\widetilde{M}_y|^2 \cos^4\theta + |\widetilde{M}_z|^2 \sin^2\theta \cos^2\theta \right. \\ &&\left. - (\widetilde{M}_y \widetilde{M}_z^{\ast} + \widetilde{M}_y^{\ast} \widetilde{M}_z) \sin\theta \cos^3\theta - i \chi \right), \nonumber \\
\frac{d \Sigma^{-+}_{\mathrm{sf}}}{d \Omega} &=& \frac{8 \pi^3}{V} b_{\mathrm{H}}^2 \left( |\widetilde{M}_x|^2 + |\widetilde{M}_y|^2 \cos^4\theta + |\widetilde{M}_z|^2 \sin^2\theta \cos^2\theta \right. \\ &&\left. - (\widetilde{M}_y \widetilde{M}_z^{\ast} + \widetilde{M}_y^{\ast} \widetilde{M}_z) \sin\theta \cos^3\theta + i \chi \right) \nonumber .
 \label{eq:equation1b}
\end{eqnarray}
The superscripts ``$+$'' and ``$-$'' refer to the neutron-spin orientation (parallel or antiparallel) relative to the direction of $\mathbf{H}_0$, $V$ denotes the scattering volume, $b_{\mathrm{H}} = 2.91 \times 10^8 \, \mathrm{A}^{-1}\mathrm{m}^{-1}$ is the magnetic scattering length in the small-angle regime (the atomic magnetic form factor is approximated by $1$, since we are dealing with forward scattering), $\widetilde{\mathbf{M}}(\mathbf{q}) = \{ \widetilde{M}_x(\mathbf{q}), \widetilde{M}_y(\mathbf{q}), \widetilde{M}_z(\mathbf{q}) \}$ represents the Fourier transform of the magnetization vector field $\mathbf{M}(\mathbf{r}) = \{ M_x(\mathbf{r}), M_y(\mathbf{r}), M_z(\mathbf{r}) \}$, $\theta$ denotes the angle between $\mathbf{q}$ and $\mathbf{H}_0$, the asterisk ``$*$'' marks the complex-conjugated quantity, $i^2 = -1$, and $\chi = \chi(\mathbf{q})$ is the chiral function. The latter quantity is obtained from (one-half times) the {\it difference} between the two spin-flip SANS cross sections, according to~\cite{michelsbook}:
\begin{eqnarray}
-i K \chi(\mathbf{q}) &=& \frac{1}{2} \left( \frac{d \Sigma^{+-}_{\mathrm{sf}}}{d \Omega} - \frac{d \Sigma^{-+}_{\mathrm{sf}}}{d \Omega} \right) \\
&=& -i K \left[ (\widetilde{M}_x \widetilde{M}_y^{\ast} - \widetilde{M}_x^{\ast} \widetilde{M}_y) \cos^2\theta - (\widetilde{M}_x \widetilde{M}_z^{\ast} - \widetilde{M}_x^{\ast} \widetilde{M}_z) \sin\theta \cos\theta \right] \nonumber ,
 \label{chiral1}
\end{eqnarray}
where $K = \frac{8 \pi^3}{V} b_{\mathrm{H}}^2$. Note that the chiral function vanishes at complete magnetic saturation ($M_x^{H_0 \rightarrow \infty} = M_y^{H_0 \rightarrow \infty} = 0$). Moreover, by expressing the magnetization Fourier components $\widetilde{M}_{x,y,z}$ in terms of their real (``R'') and imaginary (``I'') parts, i.e., $\widetilde{M}_x = \widetilde{M}_x^{\mathrm{R}} + i \widetilde{M}_x^{\mathrm{I}}$, $\widetilde{M}_x^{\ast} = \widetilde{M}_x^{\mathrm{R}} - i \widetilde{M}_x^{\mathrm{I}}$ (with $\widetilde{M}_x^{\mathrm{R}} \in \mathbb{R}$ and $\widetilde{M}_x^{\mathrm{I}} \in \mathbb{R}$) and so on for the other two components, one can rewrite $-i K \chi(\mathbf{q}) \in \mathbb{R}$ as follows:
\begin{equation}
-i K \chi(\mathbf{q}) = - 2 K \left[ (\widetilde{M}_x^{\mathrm{R}} \widetilde{M}_y^{\mathrm{I}} - \widetilde{M}_x^{\mathrm{I}} \widetilde{M}_y^{\mathrm{R}}) \cos^2\theta - (\widetilde{M}_x^{\mathrm{R}} \widetilde{M}_z^{\mathrm{I}} - \widetilde{M}_x^{\mathrm{I}} \widetilde{M}_z^{\mathrm{R}}) \sin\theta \cos\theta \right] , 
 \label{chiral2}
\end{equation}
which demonstrates that the chiral function vanishes for purely real-valued or for purely imaginary magnetization Fourier components $\widetilde{M}_{x,y,z}$. By exploiting the fact that the magnetization vector is a real-valued quantity, i.e., $M_{x,y,z}(\mathbf{r}) \in \mathbb{R}$, one can use the well-known result that the real parts of the $\widetilde{M}_{x,y,z}(\mathbf{q})$ are even functions of $\mathbf{q}$ while the imaginary parts are odd functions of $\mathbf{q}$, i.e., $\widetilde{M}_{x,y,z}^{\mathrm{R}}(\mathbf{q}) = \widetilde{M}_{x,y,z}^{\mathrm{R}}(-\mathbf{q})$ and $\widetilde{M}_{x,y,z}^{\mathrm{I}}(\mathbf{q}) = - \widetilde{M}_{x,y,z}^{\mathrm{I}}(-\mathbf{q})$. This implies that both terms in Eq.~(\ref{chiral2}), which always involve the product of two even functions (e.g., $\widetilde{M}_x^{\mathrm{R}}$ and $\sin\theta \cos\theta$) and one odd function (e.g., $\widetilde{M}_z^{\mathrm{I}}$) function, are odd functions of $\mathbf{q}$, such that the following symmetry relation holds (odd under spatial inversion of $\mathbf{q}$):
\begin{equation}
i K \chi(\mathbf{q}) = - i K \chi(-\mathbf{q}) .
 \label{chiral3}
\end{equation}
Table~\ref{tab:table1} lists the chiral function (zero or nonzero) for all the possible combinations of symmetry properties (odd or even) of the real-space magnetization components. We also refer to the review by Maleev~\cite{maleyev2002} for a discussion of the symmetry properties of the chiral function.
\begin{table}
\centering
\caption{\label{tab:table1} Summary of all the possible combinations of symmetry properties (even or odd) of the real-space magnetization components $M_{x,y,z}(\mathbf{r})$ and the ensuing symmetries (real or imaginary) of the Fourier-space magnetization components $\widetilde{M}_{x,y,z}(\mathbf{q})$ and the chiral function $\chi$ (zero or nonzero). The case that the $M_{x,y,z}(\mathbf{r})$ are composed of a nonzero even and odd part will always result in a nonzero chiral function.\\}
\small
\begin{tabular}{@{}c|c|c||c|c|c||c}
$M_x(\mathbf{r})$
&
$M_y(\mathbf{r})$
&
$M_z(\mathbf{r})$
&
$\widetilde{M}_x(\mathbf{q})$
&
$\widetilde{M}_y(\mathbf{q})$
&
$\widetilde{M}_z(\mathbf{q})$
&
$-i K \chi(\mathbf{q})$
\\
\hline
\hline
odd & odd & odd & imaginary & imaginary & imaginary & zero \\
even & odd & odd & real & imaginary & imaginary & nonzero \\
odd & even & odd & imaginary & real & imaginary & nonzero \\
even & even & odd & real & real & imaginary & nonzero \\
odd & odd & even & imaginary & imaginary & real & nonzero \\
even & odd & even & real & imaginary & real & nonzero \\
odd & even & even & imaginary & real & real & nonzero \\
even & even & even & real & real & real & zero \\
\end{tabular}
\end{table}

Besides the difference between $d \Sigma^{+-}_{\mathrm{sf}} / d \Omega$ and $d \Sigma^{-+}_{\mathrm{sf}} / d \Omega$, we can also consider (one-half times) their {\it sum}:
\begin{eqnarray}
\label{chiral444}
\frac{d \Sigma_{\mathrm{sf}}}{d \Omega} &=& \frac{1}{2} \left( \frac{d \Sigma^{+-}_{\mathrm{sf}}}{d \Omega} + \frac{d \Sigma^{-+}_{\mathrm{sf}}}{d \Omega} \right) \\
&=& K \left( |\widetilde{M}_x|^2 + |\widetilde{M}_y|^2 \cos^4\theta + |\widetilde{M}_z|^2 \sin^2\theta \cos^2\theta \nonumber \right. \\ &&\left. - (\widetilde{M}_y \widetilde{M}_z^{\ast} + \widetilde{M}_y^{\ast} \widetilde{M}_z) \sin\theta \cos^3\theta \right) . \nonumber
\end{eqnarray}
In the following, for simplicity, the quantity $d \Sigma_{\mathrm{sf}} / d \Omega$ is called the (polarization-independent) spin-flip SANS cross section. In contrast to the chiral function, $d \Sigma_{\mathrm{sf}} / d \Omega$ has the well-known property that it is an even function of $\mathbf{q}$~\cite{squires} (even under spatial inversion of $\mathbf{q}$),
\begin{equation}
\frac{d \Sigma_{\mathrm{sf}}}{d \Omega}(\mathbf{q}) = \frac{d \Sigma_{\mathrm{sf}}}{d \Omega}(-\mathbf{q}) .
\label{chiral4}
\end{equation}
Note that the cross term in Eq.~(\ref{chiral444}) can be written as $- (\widetilde{M}_y \widetilde{M}_z^{\ast} + \widetilde{M}_y^{\ast} \widetilde{M}_z) = - 2 (\widetilde{M}_y^{\mathrm{R}} \widetilde{M}_z^{\mathrm{R}} + \widetilde{M}_y^{\mathrm{I}} \widetilde{M}_z^{\mathrm{I}})$, which is an even function of $\mathbf{q}$, as are the $|\widetilde{M}_x|^2$, $|\widetilde{M}_y|^2$, and $|\widetilde{M}_z|^2$.

It is often convenient to average two-dimensional SANS data $f(\mathbf{q}) = f(q_y, q_z) = f(q, \theta)$, where $f$ either stands for $d \Sigma_{\mathrm{sf}} / d \Omega$ or for $-i K \chi$, along certain directions in $\mathbf{q}$~space, e.g.\ parallel ($\theta = 0$) or perpendicular ($\theta = \pi/2$) to the applied magnetic field, or even over the full angular $\theta$~range. In the following, we consider $2\pi$~azimuthally-averaged SANS data
\begin{equation}
\label{aziaverage}
I_{\mathrm{sf}}(q) = \frac{1}{2\pi} \int_0^{2\pi} f(q,\theta) \, d\theta ,
\end{equation}
which allows for the computation of the pair-distance distribution function $p_{\mathrm{sf}}(r)$ according to
\begin{equation}
\label{pvonreqintegral}
p_{\mathrm{sf}}(r) = r \int\limits_0^{\infty} I_{\mathrm{sf}}(q) \sin(qr) q dq .
\end{equation}
This Fourier transform corresponds to the distribution of real-space distances between volume elements inside the particle weighted by the excess scattering-length density distribution; see the reviews by Glatter~\cite{glatterchapter} and by Svergun and Koch~\cite{svergun03} for detailed discussions of the properties of $p_{\mathrm{sf}}(r)$. As a reference for nonuniformly magnetized spherical particles, we specify here the $p_{\mathrm{sf}}(r)$ of a uniformly magnetized sphere, which for $r \leq L = 2R$ equals:
\begin{equation}
\label{pvonreq}
p_{\mathrm{sf}}(r) \propto r^2 \left( 1 - \frac{3r}{4R} + \frac{r^3}{16 R^3} \right) .
\end{equation}
For the calculation of the spin-flip SANS cross section $d\Sigma_{\mathrm{sf}} / d\Omega$ [Eq.~(\ref{chiral444})] and the chiral function $-i K \chi$ [Eq.~(\ref{chiral1})], it is necessary to compute the discrete Fourier transform of all the ${\mathbf{m}}_i = {\mathbf{m}}_i({\mathbf{r}})$ belonging to the spherical nanomagnet. Using $\boldsymbol{\mu}_i = \boldsymbol{\mu}_i(\mathbf{r}) = M_{\mathrm{s}} V_i \mathbf{m}_i(\mathbf{r})$, the discrete-space Fourier transform is computed as ($V_i = a^3$):
\begin{equation}
\widetilde{\mathbf{M}}(\mathbf{q}) \cong \frac{M_{\mathrm{s}} a^3 h(\mathbf{q})}{(2\pi)^{3/2}} \sum_{i=1}^{\mathcal{K}} \mathbf{m}_i \exp\left(-\mathrm{i} \mathbf{q} \cdot \mathbf{r}_i \right) ,
\label{discreteFT}
\end{equation}
where $\mathbf{r}_i$ is the location point of the $i$th spin and $\mathbf{q}$ represents the wave vector (scattering vector). The function $h(\mathbf{q}) = \frac{\sin(q_x a/2)}{q_x a/2} \frac{\sin(q_y a/2)}{q_y a/2} \frac{\sin(q_z a/2)}{q_z a/2}$ denotes the form factor of the cubic discretization cell with $a = 2 \, \mathrm{nm}$ being the cell size; for $|q_{x,y,z}| a/2 \ll 1$, $h \rightarrow 1$. For atomistic calculations~\cite{adamsjacnum2022,adamsprb2024}, this correction is irrelevant in the small-angle regime, but for the present calculation the cell size becomes already noticeable for $q \gtrsim 0.3 \, \mathrm{nm}^{-1}$ (compare Fig.~\ref{app_6} in the Appendix). Equation~(\ref{discreteFT}) establishes the relation between the outcome of the simulations, $\mathbf{m}_i$, and $d\Sigma_{\mathrm{sf}} / d\Omega$ and $-i K \chi$. The Fourier components are evaluated in the plane $q_x=0$ (corresponding to the scattering geometry shown in Fig.~\ref{fig2} with $\mathbf{q} \cong \{ 0, q_y, q_z \} = q \{ 0, \sin\theta, \cos\theta \}$) and used in Eqs.~(\ref{chiral1}) and (\ref{chiral444}) to compute the spin-flip SANS cross section and the chiral function according to:
\begin{equation}
\left\langle f \right\rangle_{\mathrm{EA}} = \sum_{i=1}^{\mathcal{N}} f_i ,
\label{sigmaaverage}
\end{equation}
where $f_i$ represents (for fixed $H_0$) either $d \Sigma_{\mathrm{sf}} / d \Omega$ or $-i K \chi$ of a spherical particle with diameter $L$ and with a particular random easy-axis (``EA'') orientation ``$i$''. In our paper, we consider results for the SANS observables for the case of a random distribution of the cubic magnetocrystalline anisotropy axes of the particles with respect to the global direction of the external field $\mathbf{H}_0 \parallel \mathbf{e}_z$. At each value of $H_0$, micromagnetic simulations for typically $\mathcal{N} \sim 500$ random orientations between the easy particle axis and $\mathbf{H}_0$ were carried out. Equation~(\ref{sigmaaverage}) implies that interparticle-interference effects are ignored in the simulations.

\section{Results and Discussion}
\label{ressec}

\begin{figure}[tb!]
\centering
\resizebox{0.90\columnwidth}{!}{\includegraphics{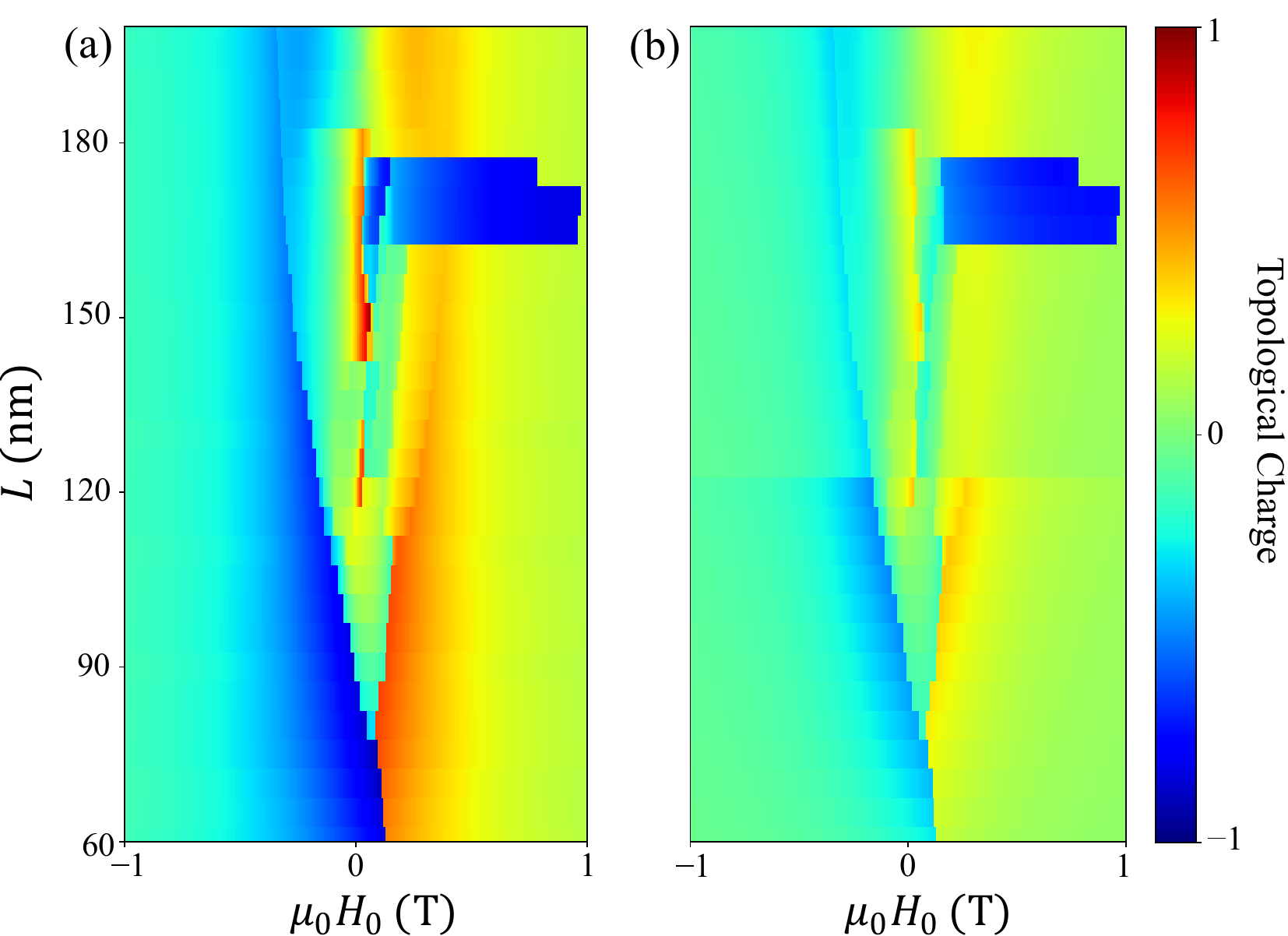}}
\caption{The field-diameter phase diagram of an oriented FeGe sphere with $L$ ranging between $60 \, \mathrm{nm}$ and $200 \, \mathrm{nm}$ and $-1 \, \mathrm{T} \leq \mu_0 H_0 \leq 1 \, \mathrm{T}$. One of the cubic anisotropy axes is parallel to the externally applied magnetic field $\mathbf{H}_0 \parallel \mathbf{e}_z$. (a)~Toplogical charge $Q$ [Eq.~(\ref{topocharge:eq})] numerically calculated for the middle layer in the $x$-$y$~plane. (b)~$Q$ averaged of all layers in the particle. The spacing (resolution) in $L$ and $H_0$ is, respectively, $5 \, \mathrm{nm}$ and $5 \, \mathrm{mT}$.}
\label{fig3}
\end{figure}

Figure~\ref{fig3} depicts the numerically-computed values of the topological charge $Q$ [Eq.~(\ref{topocharge:eq})] of a single nanoparticle for sphere diameters $L$ between $60 \, \mathrm{nm}$ and $200 \, \mathrm{nm}$ and for applied fields $\mu_0 H_0$ ranging from $-1 \, \mathrm{T}$ to $+1 \, \mathrm{T}$. In this particular example one of the cubic anisotropy axes has been chosen to be parallel to $\mathbf{H}_0$, so that the results in Fig.~\ref{fig3} are representative for an oriented particle, and not for an ensemble of randomly-oriented nanoparticles (to be discussed later). The topological charge has been computed for the middle layer in the $x$-$y$~plane [Fig.~\ref{fig3}(a)], and $Q$ has also been averaged over all the layers in the particle [Fig.~\ref{fig3}(b)]. As can be seen, the averaging procedure results (as expected) in a smearing of the data, leaving however the main features unaltered. Several regions with $Q$~values approaching unity are found indicating a possible skyrmion phase, most prominently is a region $165 \, \mathrm{nm} \lesssim L \lesssim 175 \, \mathrm{nm}$ and $0.05 \, \mathrm{T} \lesssim \mu_0 H_0 \lesssim 0.65 \, \mathrm{T}$ where $Q \rightarrow -1$.

\begin{figure}[tb!]
\centering
\resizebox{1.0\columnwidth}{!}{\includegraphics{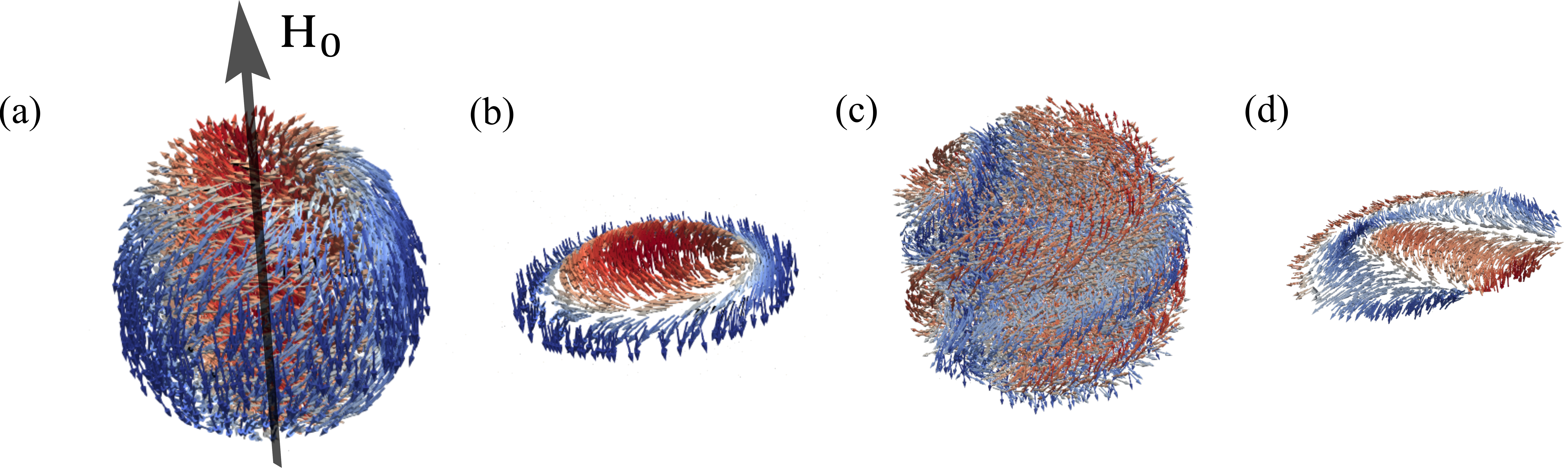}}
\caption{(a)~Spin structure (snapshot) of a $170 \, \mathrm{nm}$-sized FeGe sphere at an external magnetic field of $\mu_0 H_0 = 5 \, \mathrm{mT}$. Initially, the spin structure was saturated along $\mathbf{H}_0 \parallel \mathbf{e}_z$. One  of the three cubic anisotropy axes was chosen to be parallel to the global $\mathbf{H}_0$~direction. In panel~(b) we display the spin structure within the middle-layer $x$-$y$~plane, resulting in a topological charge of $Q \cong -0.94$. (c)~Similar to (a), but with the same cubic anisotropy axis from (a) oriented at an angle of $72^{\circ}$ relative to $\mathbf{H}_0$. (d)~Same as in (b), but with the cubic axis at $72^{\circ}$ relative to $\mathbf{H}_0$ ($Q \cong -0.39$).}
\label{fig4}
\end{figure}

The purpose of Fig.~\ref{fig3} is to demonstrate that also skyrmionic spin structures may form in individual, favorably-oriented nanoparticles of an ensemble~\cite{hertel2021}. Changing the direction of the magnetic anisotropy axes of the particle relative to the global direction given by $\mathbf{H}_0$, as it is required for the description of a particle ensemble (the subject of the paper), alters the energetics of the problem and may result in a fraction of the particles being in a skyrmion state while other particles exhibit non-topological spin structures, such as vortex- or spiral-type textures or even near single-domain structures. This is illustrated in Fig.~\ref{fig4}(a)$-$(d), where the spin structures of two differently-oriented FeGe spheres are shown. In Fig.~\ref{fig4}(a) we display the structure of a $170 \, \mathrm{nm}$-sized FeGe sphere at an external magnetic field of $5 \, \mathrm{mT}$ and with one of the cubic anisotropy axes aligned parallel to $\mathbf{H}_0$; Fig.~\ref{fig4}(b) features the spin distribution in the middle-layer $x$-$y$~plane, which is characterized by a topological charge of $Q \cong -0.94$. When the particle is oriented with the same cubic anisotropy axis at an angle of $\sim$$72^{\circ}$ relative to $\mathbf{H}_0$ [Fig.~\ref{fig4}(c)] a significantly different magnetization distribution is obtained, with $Q \cong -0.39$ in the middle-layer plane [Fig.~\ref{fig4}(d)]. These considerations imply that for a dilute set of randomly-arranged FeGe nanoparticles, the different spin configurations of differently oriented nanoparticles give rise to a spin-disorder-induced smearing of the SANS observables, even in the absence of a particle-size distribution. This smearing effect is of course the most pronounced at low fields [see, e.g., Fig.~\ref{fig7}(a) below].

\begin{figure}[tb!]
\centering
\resizebox{1.0\columnwidth}{!}{\includegraphics{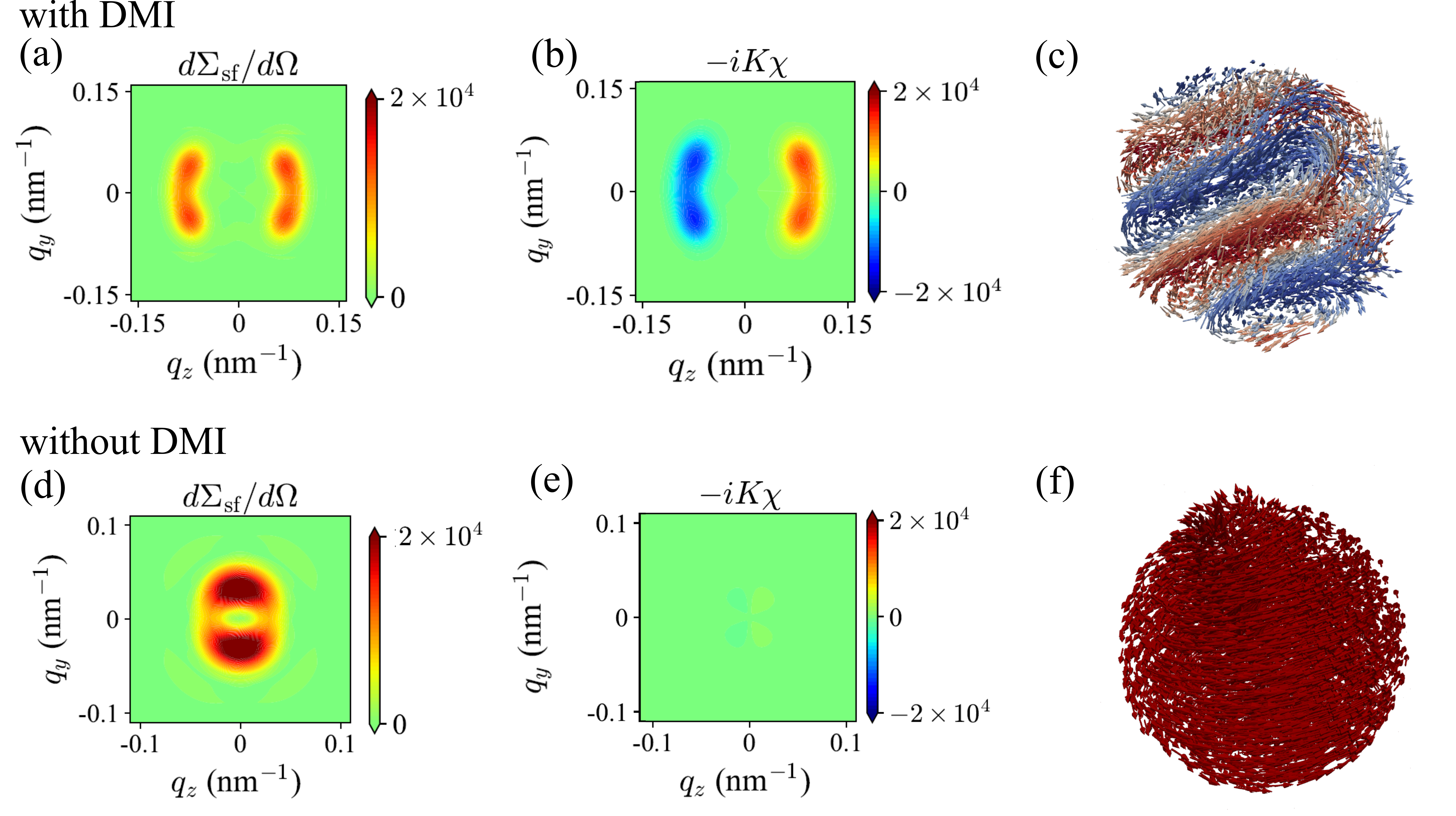}}
\caption{(a)~Spin-flip SANS cross section $d \Sigma_{\mathrm{sf}} / d \Omega$ and (b)~chiral function $-i K \chi$ (both at $5 \, \mathrm{mT}$) of an ensemble of $\mathcal{N} = 500$ randomly-oriented $170 \, \mathrm{nm}$-sized FeGe nanoparticles. (c)~Selected spin structure at $5 \, \mathrm{mT}$. (d)$-$(f)~Corresponding $d \Sigma_{\mathrm{sf}} / d \Omega$ , $-i K \chi$, and spin structure without the DMI.}
\label{fig5}
\end{figure}

The results for the randomly-averaged two-dimensional spin-flip SANS cross section $d \Sigma_{\mathrm{sf}} / d \Omega$ and for the chiral function $-i K \chi$ with and without DMI are shown in Fig.~\ref{fig5} for $\mu_0 H_0 = 5 \, \mathrm{mT}$ and $L = 170 \, \mathrm{nm}$. The Appendix features results for $d \Sigma_{\mathrm{sf}} / d \Omega$ and $-i K \chi$ for several other applied magnetic fields and diameters. Additionally, we display examples for spin structures that contribute to the respective scattering cross section [Fig.~\ref{fig5}(c) with DMI and Fig.~\ref{fig5}(f) without DMI]. Close to saturation, we (of course) always recover the characteristic $\sin^2\theta \cos^2\theta$-type angular anisotropy of $d \Sigma_{\mathrm{sf}} / d \Omega$ [compare Eq.~(\ref{chiral444})], pointing towards a uniformly magnetized nanoparticle spin structure. Reducing the field results (for a given $L$) in the emergence of a variety of complex $\mathbf{M}(\mathbf{r})$~patterns [compare, e.g., Fig.~\ref{fig4}(a) and (c)] and in a concomitant complicated randomly-averaged $d \Sigma_{\mathrm{sf}} / d \Omega$ [Fig.~\ref{fig5}(a)]. Leaving out the DMI gives rise to a drastically changed $d \Sigma_{\mathrm{sf}} / d \Omega$ [Fig.~\ref{fig5}(d)], exhibiting (here for $5 \, \mathrm{mT}$ and $L = 170 \, \mathrm{nm}$) a $\sin^2\theta$-type anisotropy that resembles the saturated state in unpolarized SANS in the $\mathbf{H}_0 \perp \mathbf{k}_0$ scattering geometry.

A central result is that in all of our simulations on randomly-arranged particle ensembles, more specifically for $60 \, \mathrm{nm} \leq L \leq 200 \, \mathrm{nm}$ and $-1 \, \mathrm{T} \leq \mu_0 H_0 \leq 1 \, \mathrm{T}$, we find a vanishing chiral function when the DMI is excluded [compare Fig.~\ref{fig5}(b) and (e) and the corresponding data in the Appendix]. In other words, due to the absence of chirality selection, the individual Fourier cross correlations in the expression for $-i K \chi$ [Eq.~(\ref{chiral2})] add up to zero in the random average case and no DMI. Similar to previous simulations on Fe nanospheres~\cite{laura2020,evelynprb2023}, we find dipolar-energy-driven vortex-type structures in FeGe when the DMI is not taken into account. The ensemble of vortex configurations exhibit, on the average, an equal amount of clockwise and counterclockwise rotation senses, so that the corresponding chiral function averages to zero. This result is somehow expected (no chirality selection), and the symmetry properties of the chiral function are well known~\cite{maleyev2002}, but here we comprehensively study the signature of the DMI on the diffuse SANS cross section of an ensemble of randomly-oriented magnetic nanoparticles.

\begin{figure}[tb!]
\centering
\resizebox{0.80\columnwidth}{!}{\includegraphics{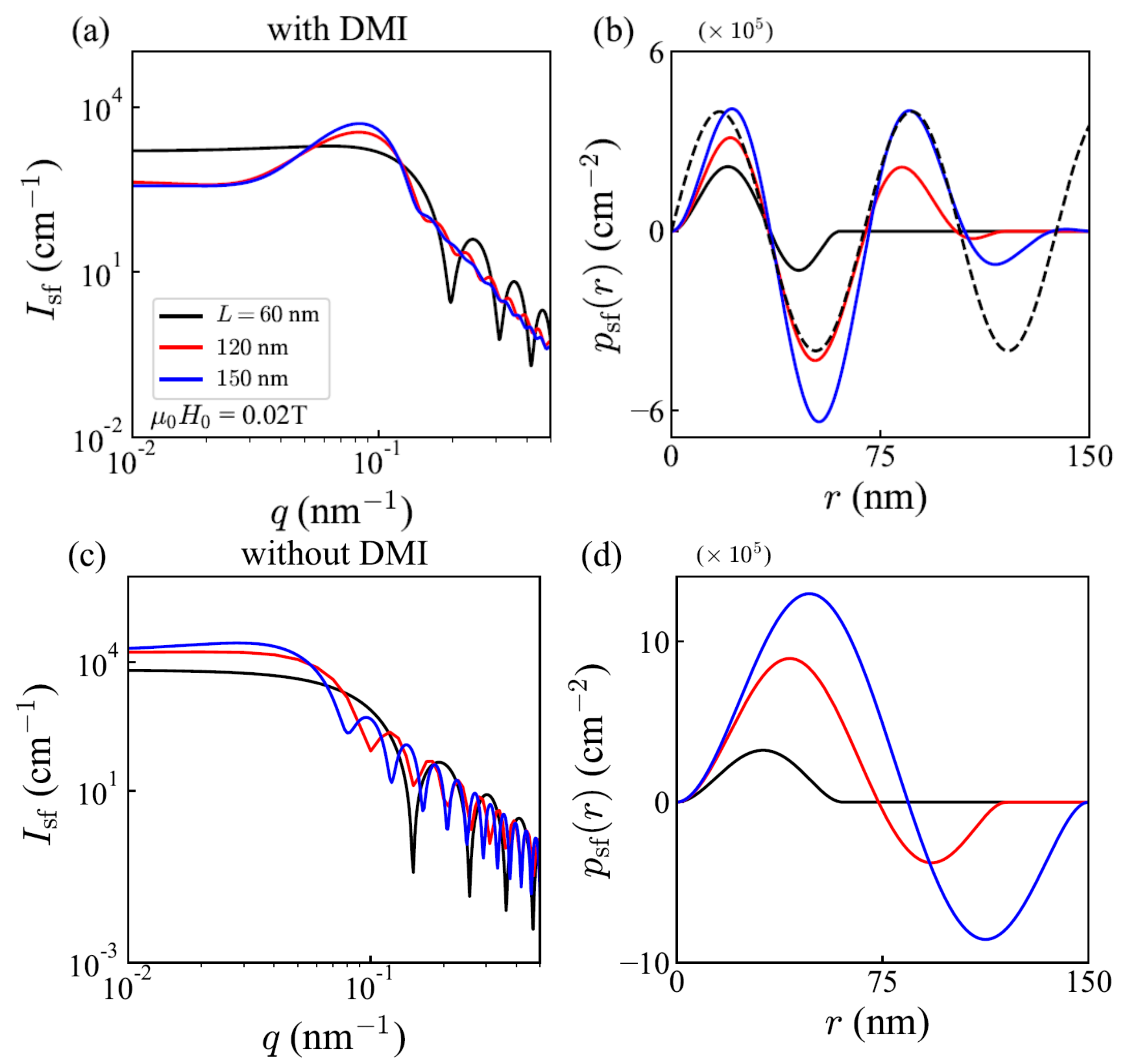}}
\caption{(a)~$I_{\mathrm{sf}}(q)$ and (b)~$p_{\mathrm{sf}}(r)$ for randomly-oriented FeGe nanoparticles with $L = 60 \, \mathrm{nm}$, $120 \, \mathrm{nm}$ and $150 \, \mathrm{nm}$ and at an applied magnetic field of $\mu_0 H_0 = 0.02 \, \mathrm{T}$ (see inset). (c)~$I_{\mathrm{sf}}(q)$ and (d)~$p_{\mathrm{sf}}(r)$ without DMI. Black dashed line in (b): $p_{\mathrm{sf}}(r) \propto \sin(k_{\mathrm{d}} r)$ with $k_{\mathrm{d}} = 0.09 \, \mathrm{nm}^{-1}$.}
\label{fig6}
\end{figure}

\begin{figure}[tb!]
\centering
\resizebox{0.85\columnwidth}{!}{\includegraphics{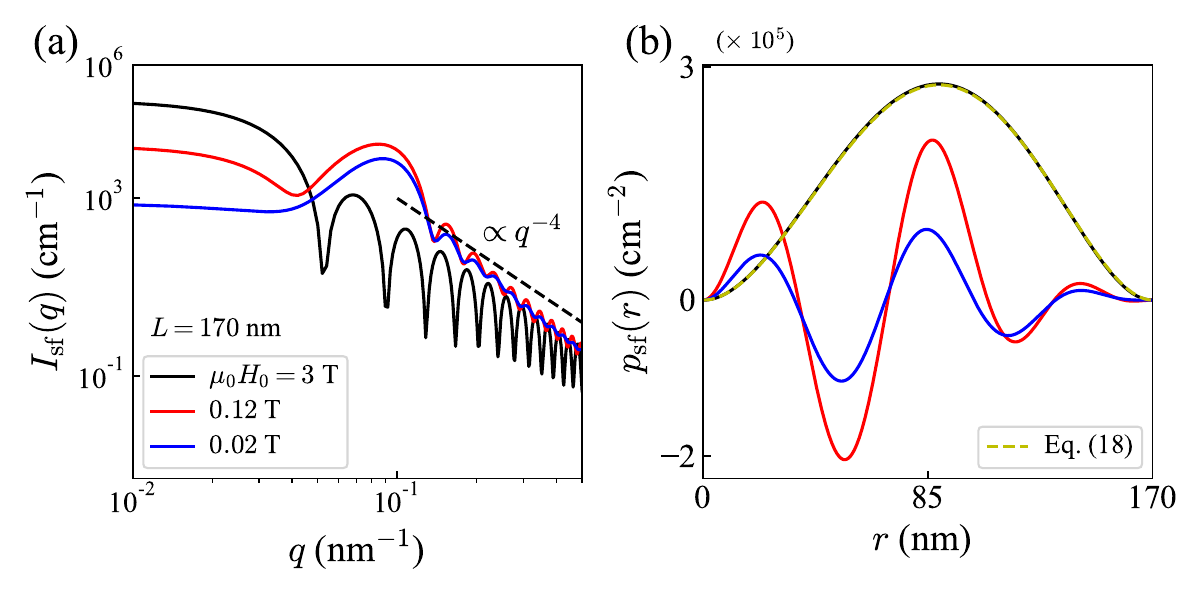}}
\caption{(a)~$I_{\mathrm{sf}}(q)$ and (b)~$p_{\mathrm{sf}}(r)$ of randomly-oriented FeGe nanoparticles with $L = 170 \, \mathrm{nm}$ and at three different applied magnetic fields ($3 \, \mathrm{T}$, $0.12 \, \mathrm{T}$, $0.02 \, \mathrm{T}$, see inset). The DMI is included in the simulations. Black dashed line in (a):~$I_{\mathrm{sf}}(q) \propto q^{-4}$. Yellow dashed line in (b):~analytical expression for a uniformly magnetized sphere [Eq.~(\ref{pvonreq})].} 
\label{fig7}
\end{figure}

The results for the azimuthally-averaged neutron data $I_{\mathrm{sf}}(q)$ along with the pair-distance distribution $p_{\mathrm{sf}}(r)$ are displayed in Figs.~\ref{fig6} and \ref{fig7}. Figure~\ref{fig6} shows the effect of the DMI for FeGe particle sizes of $L = 60 \, \mathrm{nm}$, $120 \, \mathrm{nm}$ and $150 \, \mathrm{nm}$ and at an applied magnetic field of $\mu_0 H_0 = 0.02 \, \mathrm{T}$, while Fig.~\ref{fig7} highlights the field dependence of $I_{\mathrm{sf}}(q)$ and $p_{\mathrm{sf}}(r)$ at a fixed particle size of $L = 170 \, \mathrm{nm}$ (including the DMI). Although it is difficult to make general statements regarding the spin structure of individual nanoparticles, we observe the tendency of the formation of periodic domain structures when the DMI is included [see, e.g., Fig.~\ref{fig5}(c))]. This can be seen in the $p_{\mathrm{sf}}(r)$~data, which (for $L = 120 \, \mathrm{nm}$ and $L = 150 \, \mathrm{nm}$) exhibit three zero crossings with DMI [Fig.~\ref{fig6}(b)], while only one such zero crossing is seen when the DMI is excluded [Fig.~\ref{fig6}(d)]. The observation of only one such zero crossing in $p_{\mathrm{sf}}(r)$ is indicative of a vortex-type spin structure~\cite{laura2020}. The $L = 60 \, \mathrm{nm}$ spheres are in a nearly single-domain state without DMI [Fig.~\ref{fig6}(d)], and reveal a vortex-type spin structure with DMI [Fig.~\ref{fig6}(b)]. For particle sizes that are roughly larger than the single-domain limit ($l_{\mathrm{sd}} \cong 72 \sqrt{A K_{\mathrm{c}1}}/(\mu_0 M_{\mathrm{s}}^2) = 115 \, \mathrm{nm}$), the ``periodicity'' of the $p_{\mathrm{sf}}(r)$ in Fig.~\ref{fig6}(b) (dashed line) can be well reproduced by the following characteristic wave number $k_{\mathrm{d}} = 2\pi / l_{\mathrm{d}} \cong 0.09 \, \mathrm{nm}^{-1}$ with $l_{\mathrm{d}} = 4\pi A/D = 69.1 \, \mathrm{nm}$ being the helical period.

As becomes visible in Fig.~\ref{fig7}, at a saturating field of $3 \, \mathrm{T}$, we recover the results for a homogeneously magnetized sphere [yellow dashed line in Fig.~\ref{fig7}(b)]. Reducing the field to $0.12 \, \mathrm{T}$ and $0.02 \, \mathrm{T}$ results in the already mentioned spin-disorder-induced smearing of the scattering curves [Fig.~\ref{fig7}(a)]; the form factor oscillations (most prominent at $3 \, \mathrm{T}$) get smeared and damped. Asymptotically, at large $q$, where structure on a real-space length scale of only a few nanometers is probed, we find for all scattering curves the familiar Porod law $I_{\mathrm{sf}}(q) \propto q^{-4}$ [see blacked dashed line in Fig.~\ref{fig7}(a)]. Despite the possible highly inhomogeneous internal spin structure, the asymptotic behavior of $I_{\mathrm{sf}}(q)$ is determined by the discontinuous jump of the magnetization at the particle surface, which results in the $q^{-4}$~dependency. We also see that the nucleation of an inhomogeneous spin structure at lower fields is accompanied by the formation of a maximum in the $I_{\mathrm{sf}}(q)$~curve at intermediate momentum transfers and the concomitant reduction of the value of $I_{\mathrm{sf}}(q)$ when $q \rightarrow 0$. The latter observation is due to the fact that the value of the spin-flip SANS cross section at $q=0$ reflects the behavior of the average ensemble magnetization, which decreases with decreasing field [compare to Eqs.~(\ref{chiral444}) and (\ref{discreteFT})].

To describe the scattering behavior of the randomly-averaged system at low fields and for not too small particle sizes (so that a DMI-induced spin modulation appears), we introduce the following expression for the correlation function:
\begin{eqnarray}
c_{\mathrm{sf}}(r) = B \, j_0(k_{\mathrm{d}} r) \exp(-r/R)
\label{cvonranalytical},
\end{eqnarray}
where $B$ is a scaling constant, $R = L/2$ is the sphere radius, and $j_0(z) = \sin(z)/z$ denotes the zeroth-order spherical Bessel function that provides a damped oscillation with a wave number of $\sim$$k_{\mathrm{d}}$. We emphasize that Eq.~(\ref{cvonranalytical}) does not represent a true particle correlation function, since it extends to infinity and vanishes for $r > L$. The exponential decay forces the spatial extent of $c_{\mathrm{sf}}(r)$ to be roughly limited to $r \lesssim L$. Equation~(\ref{cvonranalytical}) is an easy-to-implement expression that, as we will see below, grasps the main low-field characteristics found in the simulations. The corresponding analytical expression for the spin-flip SANS cross section reads:
\begin{eqnarray}
\label{ivonqanalytical}
I_{\mathrm{sf}}(q) &=& \int\limits_0^{\infty} c_{\mathrm{sf}}(r) j_0(qr) r^2 dr \\
 &=& \frac{2 B R^3}{\left[ 1 + \left( q - k_{\mathrm{d}} \right)^2 R^2 \right] \left[ 1 + \left( q + k_{\mathrm{d}} \right)^2 R^2 \right]} \nonumber , 
\end{eqnarray}
which exhibits a field-independent maximum at $q_{\mathrm{max}} = \sqrt{ k_{\mathrm{d}}^2 R^2 - 1} / (R) \cong k_{\mathrm{d}}$ and an asymptotic $q^{-4}$~dependency; $I_{\mathrm{sf}}(q=0) = 2 B R^3/(1+k_{\mathrm{d}}^2 R^2)^2$.

\begin{figure}[tb!]
\centering
\resizebox{1.0\columnwidth}{!}{\includegraphics{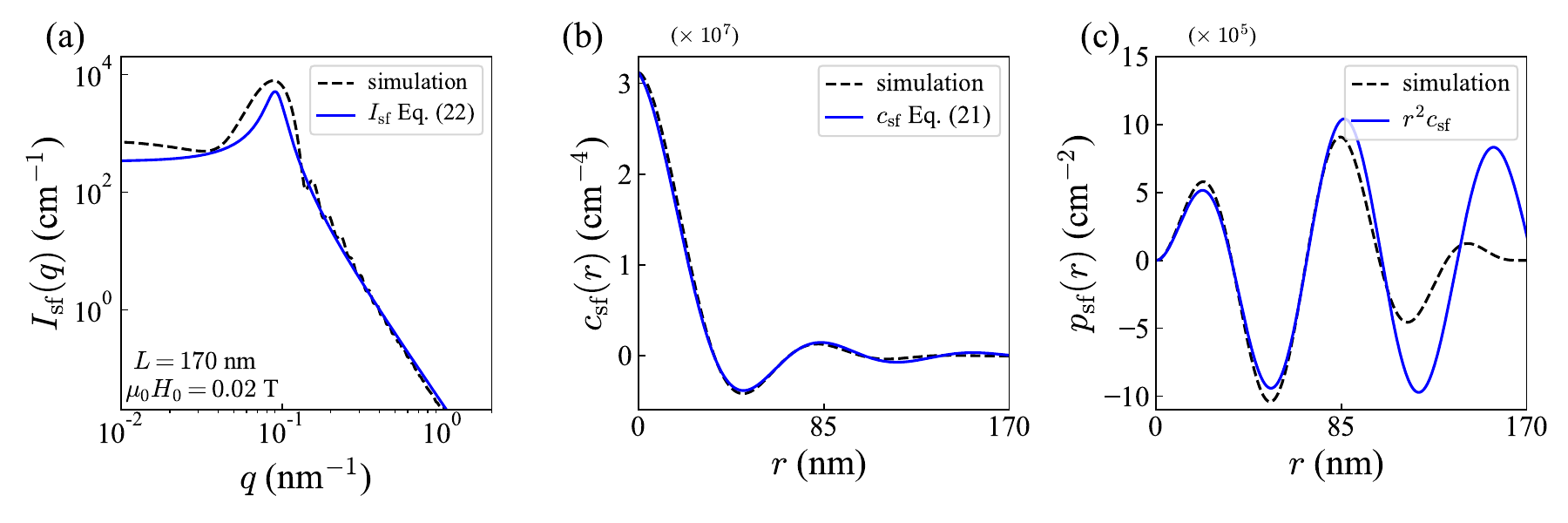}}
\caption{Comparison between the simplified analytical model [Eqs.~(\ref{cvonranalytical}) and (\ref{ivonqanalytical})] and the numerical micromagnetic simulations. (a)~$I_{\mathrm{sf}}(q)$, (b)~$c_{\mathrm{sf}}(r)$, and (c)~$p_{\mathrm{sf}}(r)$ of randomly-oriented FeGe nanoparticles ($L = 170 \, \mathrm{nm}$ and $\mu_0 H_0 = 0.02 \, \mathrm{T}$). Black dashed lines:~micromagnetic simulation. Blue solid lines:~analytical model (scaled to the simulation data).} 
\label{fig8}
\end{figure}

Figure~\ref{fig8} features a comparison between Eqs.~(\ref{cvonranalytical}) and (\ref{ivonqanalytical}) and the numerically computed $I_{\mathrm{sf}}(q)$, $c_{\mathrm{sf}}(r)$, and $p_{\mathrm{sf}}(r) = r^2 c_{\mathrm{sf}}(r)$. Overall, we see that the analytical expressions reproduce the main features of the spin-flip scattering, i.e., a peak at about the helical wavevector $k_{\mathrm{d}}$ followed by a $q^{-4}$~Porod decay at large $q$. The $r^2$~factor in the definition of $p_{\mathrm{sf}}(r)$ amplifies the error at the larger distances. The behavior of $I_{\mathrm{sf}}(q)$ at large $q$ does not depend on $k_{\mathrm{d}}$.

\section{Conclusion}
\label{summary}

Using numerical micromagnetic computations, we have investigated the signature of the antisymmetric Dzyaloshinkii-Moriya interaction (DMI) in the diffuse magnetic spin-flip small-angle neutron scattering cross section (SANS) of an ensemble of randomly-oriented FeGe nanoparticles. The DMI energy is a pseudoscalar that breaks space-inversion symmetry, in contrast to the other magnetic energies that are considered in our simulations (isotropic exchange, magnetic anisotropy, dipolar and Zeeman energies). Depending on the relative orientation between the magnetic anisotropy axes of the nanoparticles and the global direction of the externally applied magnetic field, a variety of different spin structures may appear in nanoparticles of a given size class $L$ (e.g., skyrmions, vortex- and spiral-type, nearly single domain). This results in an intrinsic spin-disorder-induced broadening of the spin-flip SANS cross section (even when all the particles have the same size). Within the studied parameter space for the particle size ($60 \, \mathrm{nm} \leq L \leq 200 \, \mathrm{nm}$) and the applied magnetic field ($-1 \, \mathrm{T} \leq \mu_0 H_0 \leq 1 \, \mathrm{T}$), we find that the randomly-averaged chiral function $-iK\chi$ is only nonzero when the DMI is taken into account in the simulations. An interesting open question in this context addresses the relation between the symmetry properties of the micromagnetic energies (and conjugate fields) under space inversion and the real and imaginary parts of the $\widetilde{M}_{x,y,z}(\mathbf{q})$. For this, Brown's nonlinear equations would need to be Fourier transformed, which involves however complicated convolution products that cannot be evaluated straightforwardly. Only within a linearized analytical approach, suitable for bulk ferromagnets, it has been shown in Ref.~\cite{michelsPRB2016} that a nonzero DMI results in complex $\widetilde{M}_{x,y,z}(\mathbf{q})$ and, consequently, in the appearance of a nonzero chiral function. Motivated by the appearance of low-field spin textures that are modulated by the characteristic wave number $k_{\mathrm{d}} = D/(2A)$, we have suggested analytical expressions for the correlation function and the ensuing SANS cross section [Eqs.~(\ref{cvonranalytical}) and (\ref{ivonqanalytical})] that are able to reproduce the main features of a random ensemble of FeGe nanoparticles.

\acknowledgments{Evelyn Pratami Sinaga, Michael P.\ Adams, and Andreas Michels acknowledge financial support from the National Research Fund of Luxembourg (PRIDE MASSENA Grant and AFR Grant No.~15639149). Eddwi H.\ Hasdeo  acknowledges financial support from the National Research Fund of Luxembourg under grant C21/MS/15752388/NavSQM. The simulations presented in this paper were carried out using the HPC facilities of the University of Luxembourg (\url{https://hpc.uni.lu}). The authors thank Konstantin L.\ Metlov (Donetsk Institute for Physics and Technology) for critically reading the manuscript.}

\appendix
\section*{Overview of SANS results for the spin-flip SANS cross section and the chiral function with and without the Dzyaloshinskii-Moriya interaction}
\label{appendix}

\begin{figure}[tb!]
\centering
\resizebox{0.90\columnwidth}{!}{\includegraphics{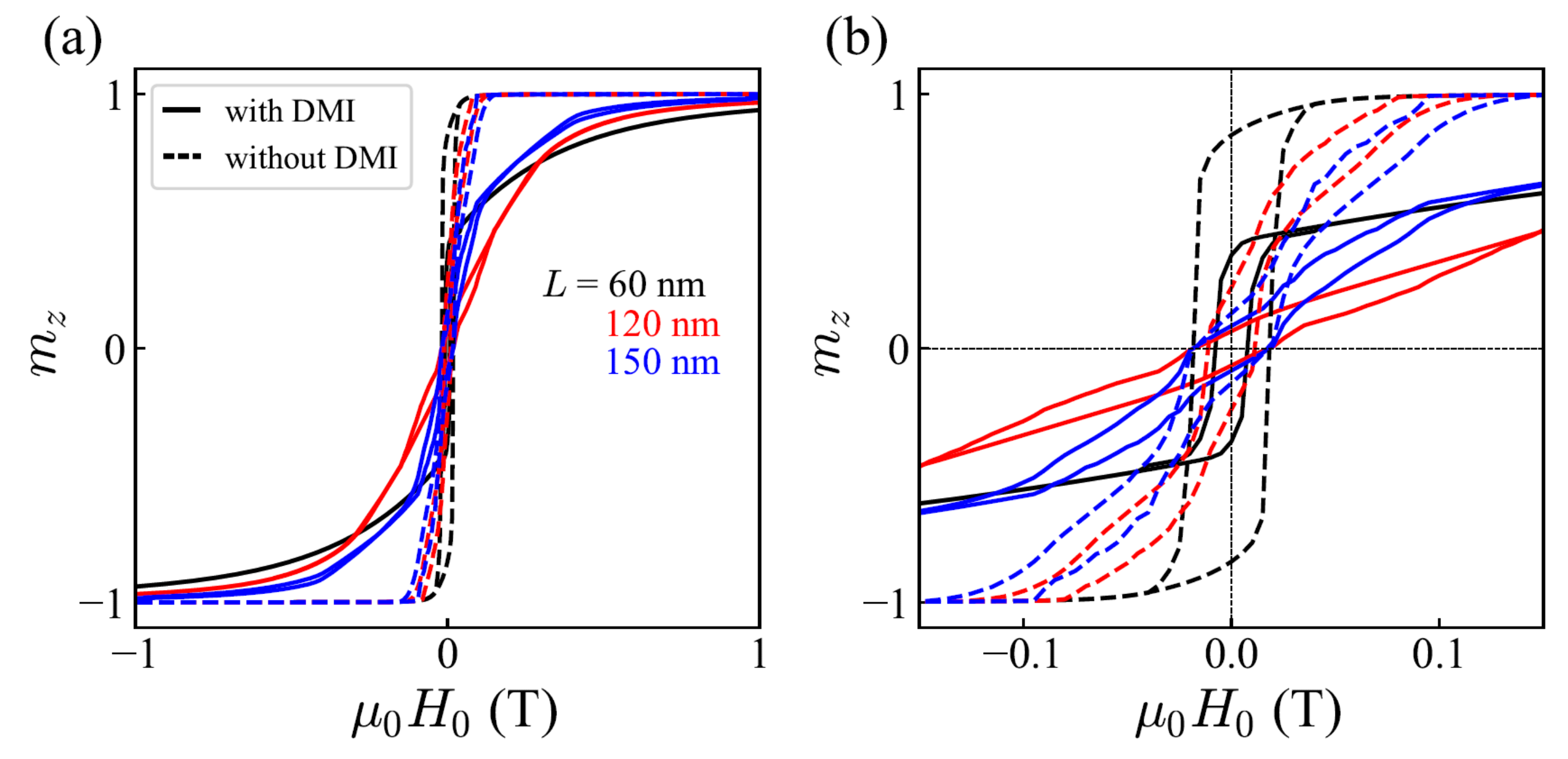}}
\caption{(a)~Normalized magnetization curves of randomly-oriented FeGe nanoparticles with particle diameters of $L = 60 \, \mathrm{nm}$, $120 \, \mathrm{nm}$, and $150 \, \mathrm{nm}$ (see inset). The solid lines are with DMI and the dashed lines are without DMI. (b)~Same as (a), but for $-0.15 \, \mathrm{T} \leq \mu_0 H_0 \leq 0.15 \, \mathrm{T}$. The reduced remanence of the $L = 60 \, \mathrm{nm}$~``sample'' is $\sim$$0.832$ (without DMI), which is very close to the Stoner-Wohlfarth value, suggesting the presence of single-domain particles~\cite{usov1997}.}
\label{app_1}
\end{figure}

\begin{figure}[tb!]
\centering
\resizebox{1.0\columnwidth}{!}{\includegraphics{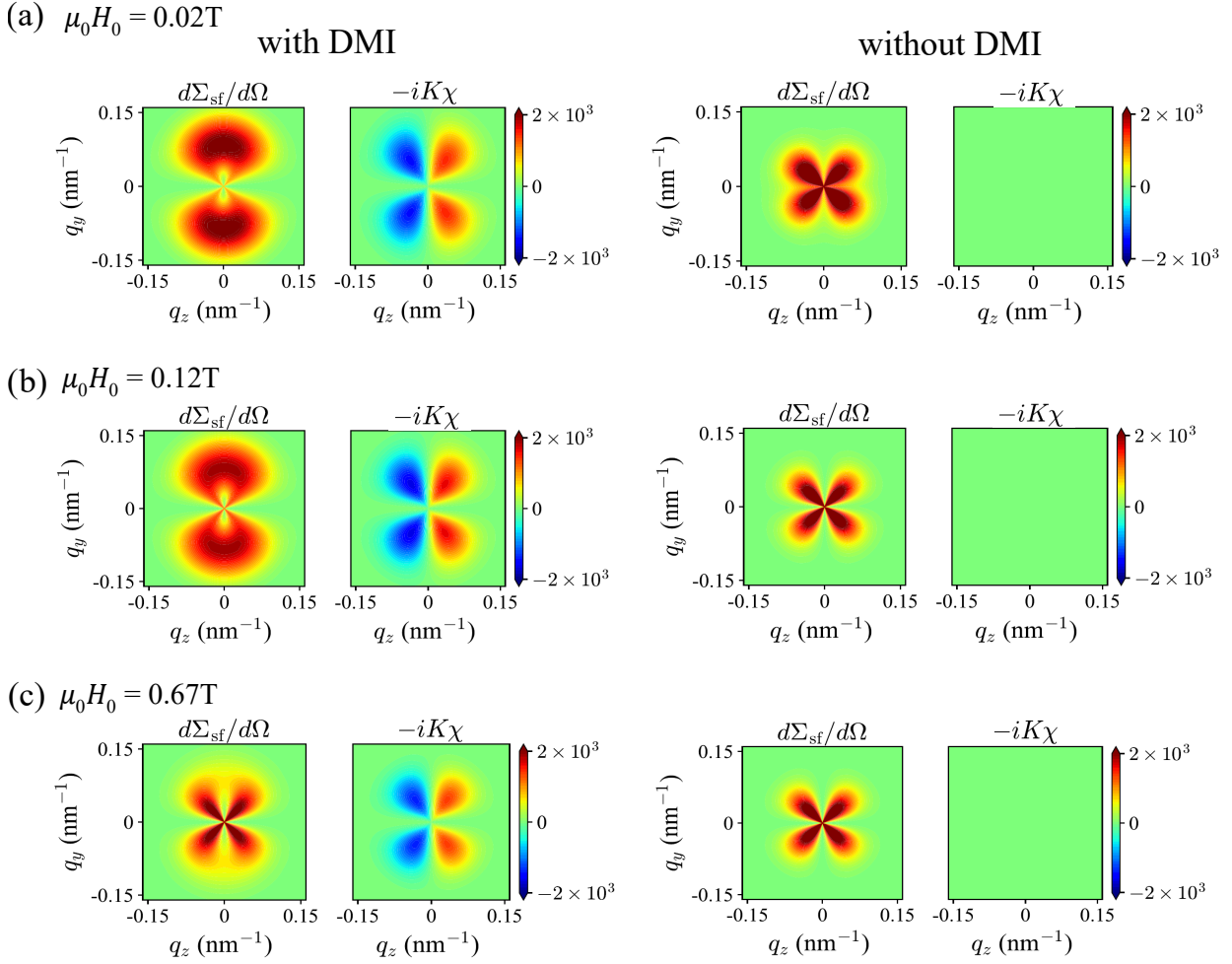}}
\caption{Comparison of simulation results for the randomly-averaged spin-flip SANS cross section $d \Sigma_{\mathrm{sf}} / d \Omega$ and the chiral function $-i K \chi$ of $L = 60 \, \mathrm{nm}$-sized FeGe nanoparticles at a series of applied magnetic fields. (a)~$\mu_0 H_0 = 0.02 \, \mathrm{T}$; (b)~$\mu_0 H_0 = 0.12 \, \mathrm{T}$; (c)~$\mu_0 H_0 = 0.67 \, \mathrm{T}$. The left panel shows simulation results with DMI, while the data in the right panel have no DMI.}
\label{app_2}
\end{figure}

In this Appendix we display additional results for the randomly-averaged magnetization curve, the spin-flip SANS cross section $d \Sigma_{\mathrm{sf}} / d \Omega$, the chiral function $-i K \chi$, and for the pair-distance distribution function $p_{\mathrm{sf}}(r)$ of FeGe nanoparticles [Figs.~\ref{app_1}$-$\ref{app_6}]. All the magnetic interactions [Eqs.~(\ref{eq1})$-$(\ref{eq5})] were taken into account in the simulations, and we compare results with and without the DMI energy. Within the scanned parameter space ($60 \, \mathrm{nm} \leq L \leq 200 \, \mathrm{nm}$ and $-1 \, \mathrm{T} \leq \mu_0 H_0 \leq 1 \, \mathrm{T}$), we find a vanishing chiral function for the case when the DMI is absent. Figure~\ref{app_1} shows that the inclusion of the DMI results in a reduced remanent magnetization of the particle ensemble, as was previously reported in Ref.~\cite{erokhinnjp2023}. Figure~\ref{app_5} highlights the scaling of the peak maximum in $I_{\mathrm{sf}}(q)$ and $p_{\mathrm{sf}}(r)$ with the DMI constant, and Fig.~\ref{app_6} shows the effect of the form factor of the cubic discretization cell, $h(\mathbf{q})$, on the randomly-averaged $I_{\mathrm{sf}}(q)$.

\begin{figure}[tb!]
\centering
\resizebox{1.0\columnwidth}{!}{\includegraphics{app_3.pdf}}
\caption{Similar to Fig.~\ref{app_2}, but for $L = 120 \, \mathrm{nm}$.}
\label{app_3}
\end{figure}

\begin{figure}[tb!]
\centering
\resizebox{1.0\columnwidth}{!}{\includegraphics{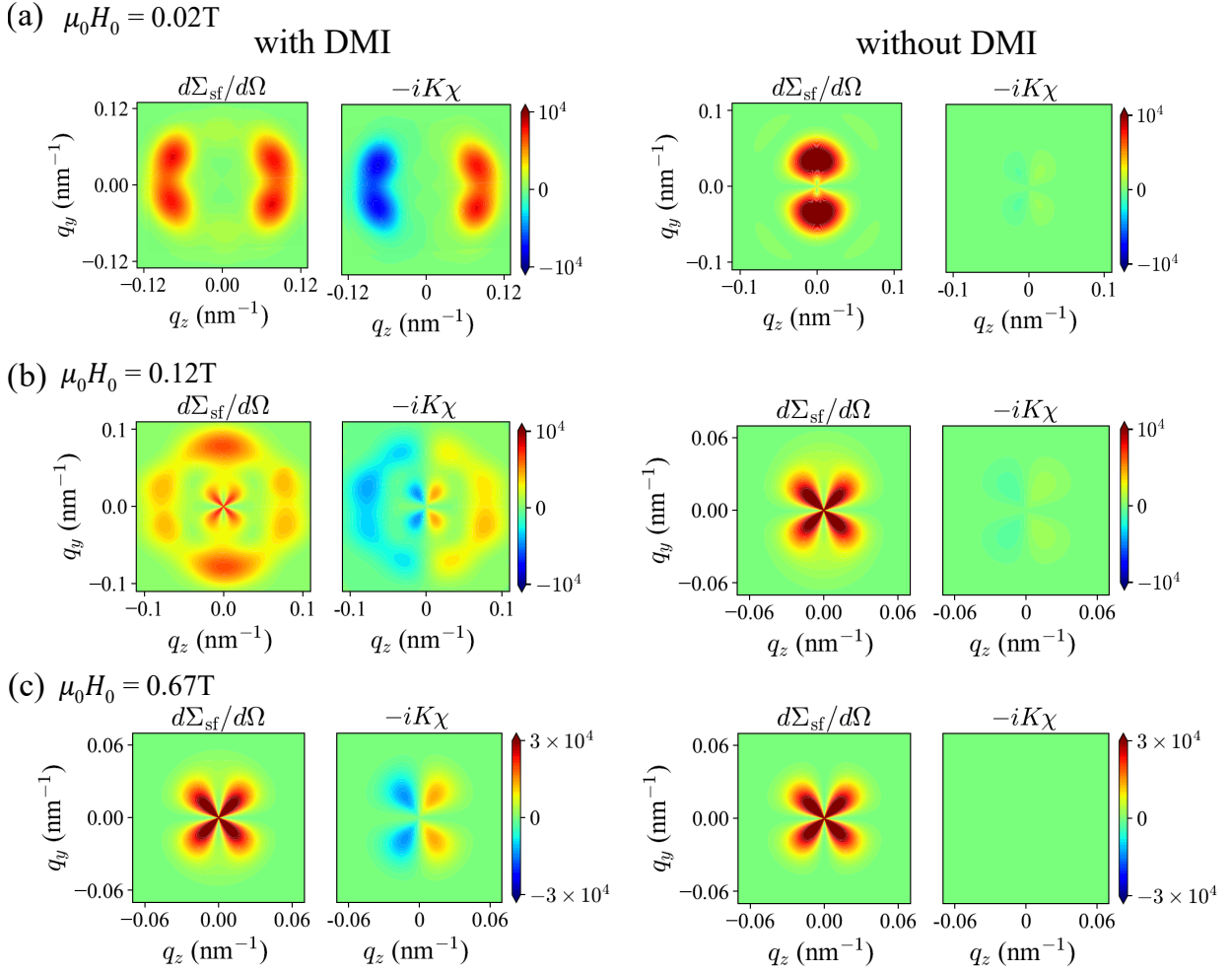}}
\caption{Similar to Fig.~\ref{app_2}, but for $L = 150 \, \mathrm{nm}$.}
\label{app_4}
\end{figure}

\begin{figure}[tb!]
\centering
\resizebox{0.90\columnwidth}{!}{\includegraphics{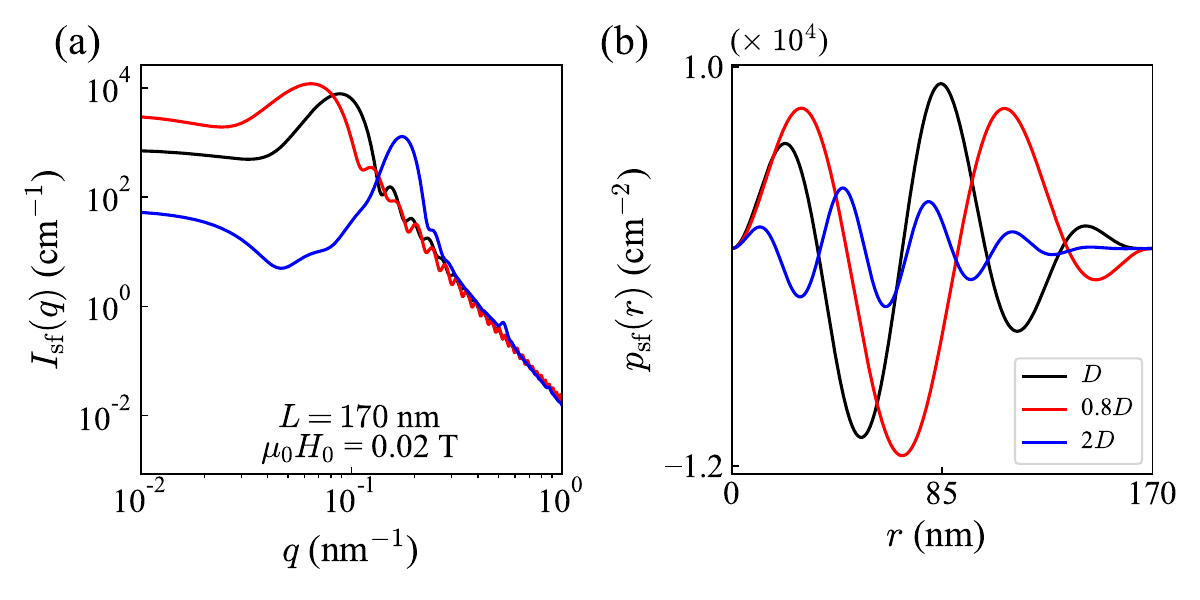}}
\caption{Dependence of the randomly-averaged $I_{\mathrm{sf}}(q)$~(a) and $p_{\mathrm{sf}}(r)$~(b) of FeGe nanospheres on the DMI constant (see inset) ($L = 170 \, \mathrm{nm}$ and $\mu_0 H_0 = 0.02 \, \mathrm{T}$). The peak maximum in $I_{\mathrm{sf}}(q)$ scales with $k_{\mathrm{d}} = D/(2A)$. Peak positions in (a): $0.062 \, \mathrm{nm}^{-1}$, $0.086 \, \mathrm{nm}^{-1}$, and $0.172 \, \mathrm{nm}^{-1}$.}
\label{app_5}
\end{figure}

\begin{figure}[tb!]
\centering
\resizebox{0.60\columnwidth}{!}{\includegraphics{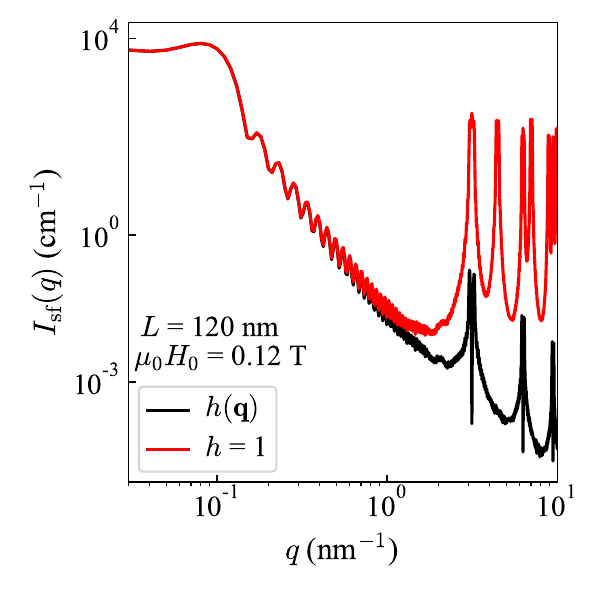}}
\caption{Effect of the form factor of the cubic discretization cell, $h(\mathbf{q})$, on the randomly-averaged spin-flip SANS cross section $I_{\mathrm{sf}}(q)$ [compare Eq.~(\ref{discreteFT})]. Shown is $I_{\mathrm{sf}}(q)$ for $L = 120 \, \mathrm{nm}$ and at $\mu_0 H_0 = 0.12 \, \mathrm{T}$ with the function $h(q_x=0,q_y,q_z) = \frac{\sin(q_y a/2)}{q_y a/2} \frac{\sin(q_z a/2)}{q_z a/2}$ included using a cell size of $a = 2 \, \mathrm{nm}$ (black line) and for $h = 1$ (red line) (log-log scale). As is seen, the $q$~dependent cell form factor suppresses the scattering curve. Here, significant deviations become noticeable for $q \gtrsim 0.3 \, \mathrm{nm}^{-1}$.}
\label{app_6}
\end{figure}

\bibliography{Pratami_et_al_2023}

\end{document}